\documentclass[twocolumn,showpacs,preprintnumbers,amsmath,prl,graphicx,amssymb,superscriptaddress]{revtex4}
\usepackage{graphicx}% Include figure files
\usepackage{dcolumn}% Align table columns on decimal point
\usepackage{bm}% bold math
\usepackage{psfrag}

\newcommand{\snoccfluxshort}{1.76^{+0.06}_{-0.05}\mbox{(stat.)}^{+0.09}_{-0.09}~\mbox{(syst.)}} 
\newcommand{\snoesfluxshort}{2.39^{+0.24}_{-0.23}\mbox{(stat.)}^{+0.12}_{-0.12}~\mbox{(syst.)}} 
\newcommand{\snoncfluxshort}{5.09^{+0.44}_{-0.43}\mbox{(stat.)}^{+0.46}_{-0.43}~\mbox{(syst.)}} 
\newcommand{\snoncfluxunc}{6.42^{+1.57}_{-1.57}\mbox{(stat.)}^{+0.55}_{-0.58}~\mbox{(syst.)}} 
\newcommand{\snomutauflux}{3.41^{+0.45}_{-0.45}\mbox{(stat.)}^{+0.48}_{-0.45}~\mbox{(syst.)}} 
\newcommand{\snoeflux}{1.76^{+0.05}_{-0.05}\mbox{(stat.)}^{+0.09}_{-0.09}~\mbox{(syst.)}} 
\newcommand{\snomutaufluxsk}{3.45^{+0.65}_{-0.62}} 
\newcommand{\snomutaufluxcomb}{3.41^{+0.66}_{-0.64}} 
\newcommand{\nccfit}{1967.7^{+61.9}_{-60.9}} 
\newcommand{\nesfit}{263.6^{+26.4}_{-25.6}} 
\newcommand{\nncfit}{576.5^{+49.5}_{-48.9}}  
\newcommand{\nsigmassno}{5.3} 
\newcommand{\nsigmassk}{5.5} 

\newcommand{\ssmflux}{5.05^{+1.01}_{-0.81}}
\newcommand{\phisk}{2.32\pm0.03\mbox{(stat.)}^{+0.08}_{-0.07}~\mbox{(syst.)}}

\renewcommand{\today}{\number\day\space\ifcase\month\or January\or 
 February\or March\or April\or May\or June\or July\or August\or 
 September\or October\or November\or December\fi\space\number\year}

\begin{document}
%%\preprint{PREPRINT No.}
\title{Direct Evidence for Neutrino Flavor Transformation from Neutral-Current Interactions in the Sudbury Neutrino Observatory}
%%Begin Author List
%Authorlist v.2.8 May 6, 2002
%
\newcommand{\ubc}{Department of Physics and Astronomy, University of 
British Columbia, Vancouver, BC V6T 1Z1 Canada}
\newcommand{\bnl}{Chemistry Department, Brookhaven National 
Laboratory,  Upton, NY 11973-5000}
\newcommand{\carleton}{Carleton University, Ottawa, Ontario K1S 5B6 Canada}
\newcommand{\uog}{Physics Department, University of Guelph,  
Guelph, Ontario N1G 2W1 Canada}
\newcommand{\lu}{Department of Physics and Astronomy, Laurentian 
University, Sudbury, Ontario P3E 2C6 Canada}
\newcommand{\lbnl}{Institute for Nuclear and Particle Astrophysics and 
Nuclear Science Division, Lawrence Berkeley National Laboratory, Berkeley, CA 94720}
\newcommand{\lanl}{Los Alamos National Laboratory, Los Alamos, NM 87545}
\newcommand{\oxford}{Department of Physics, University of Oxford, 
Denys Wilkinson Building, Keble Road, Oxford, OX1 3RH, UK}
\newcommand{\penn}{Department of Physics and Astronomy, University of 
Pennsylvania, Philadelphia, PA 19104-6396}
\newcommand{\queens}{Department of Physics, Queen's University, 
Kingston, Ontario K7L 3N6 Canada}
\newcommand{\uw}{Center for Experimental Nuclear Physics and Astrophysics, 
and Department of Physics, University of Washington, Seattle, WA 98195}
\newcommand{\triumf}{TRIUMF, 4004 Wesbrook Mall, Vancouver, BC V6T 2A3, Canada}
\newcommand{\ralsuss}{Rutherford Appleton Laboratory, Chilton, Didcot, 
Oxon, OX11 0QX, and University of Sussex, Physics and Astronomy Department, 
Brighton BN1 9QH, UK}
\newcommand{\uci}{Department of Physics, University of California, 
Irvine, CA 92717}
\newcommand{\aecl}{Atomic Energy of Canada, Limited, Chalk River Laboratories, 
Chalk River, ON K0J 1J0, Canada}
\newcommand{\nrc}{National Research Council of Canada, Ottawa, ON K1A 0R6, Canada}
\newcommand{\princeton}{Department of Physics, Princeton University, 
Princeton, NJ 08544}
\newcommand{\birkbeck}{Birkbeck College, University of London, Malet Road, 
London WC1E 7HX, UK}
%%%%%%%%%%%%

\affiliation{	\aecl	}
\affiliation{	\ubc	}
\affiliation{	\bnl	}
\affiliation{	\uci	}
\affiliation{	\carleton	}
\affiliation{	\uog	}
\affiliation{	\lu	}
\affiliation{	\lbnl	}
\affiliation{	\lanl	}
\affiliation{	\nrc	}
\affiliation{	\oxford	}
\affiliation{	\penn	}
\affiliation{	\princeton	}
\affiliation{	\queens	}
\affiliation{	\ralsuss	}
\affiliation{	\triumf	}
\affiliation{	\uw	}

\author{	Q.R.~Ahmad	}\affiliation{	\uw	}
\author{	R.C.~Allen	}\affiliation{	\uci	}
\author{	T.C.~Andersen	}\affiliation{	\uog	}			
\author{	J.D.~Anglin	}\affiliation{	\nrc	}			
\author{	J.C.~Barton	}\altaffiliation[Permanent Address: ]{\birkbeck}	\affiliation{	\oxford	}		
\author{	E.W.~Beier	}\affiliation{	\penn	}			
\author{	M.~Bercovitch	}\affiliation{	\nrc	}			
\author{	J.~Bigu	}\affiliation{	\lu	}			
\author{	S.D.~Biller	}\affiliation{	\oxford	}			
\author{	R.A.~Black	}\affiliation{	\oxford	}			
% ...

\collaboration{SNO Collaboration}
\noaffiliation
%%End Author List
\date{\today}% It is always \today, today,
\begin{abstract}
Observations of neutral-current $\nu$ interactions on deuterium in the Sudbury Neutrino Observatory
are reported.  Using the neutral current, elastic scattering, and charged current reactions and assuming the standard ${}^{8}$B shape, the $\nu_{e}$ component of the ${}^{8}$B solar flux is $\phi_{e}=\snoeflux\times 10^6~{\rm cm}^{-2} {\rm s}^{-1}$ for a kinetic energy threshold of 5 MeV.
The non-$\nu_{e}$  component is $\phi_{\mu\tau}=\snomutauflux\times 10^6~{\rm cm}^{-2} {\rm s}^{-1}$, $\nsigmassno\sigma$ greater than zero, providing  strong evidence for solar $\nu_{e}$ flavor transformation.  The total flux measured with the NC reaction is $\phi_{\text{NC}}=\snoncfluxshort\times 10^6~{\rm cm}^{-2} {\rm s}^{-1}$, consistent with solar models. 
\end{abstract}

\maketitle
The Sudbury Neutrino Observatory (SNO) detects $^8$B solar neutrinos through the reactions:
\begin{displaymath}
\begin{array}{l l l l}
\nu_e + \text{d}          & \rightarrow &
        \text{p} \!+ \text{p} + \text{e}^- & \qquad\text{(CC)}, \\
\nu_x + \text{d}          & \rightarrow &
        \text{p} + \text{n} + \nu_x        & \qquad\text{(NC)}, \\
\nu_x + \text{e}^-\!\!\!\!& \rightarrow &
        \nu_x + \text{}e^-                 & \qquad\text{(ES).}
\end{array}
\end{displaymath}
The charged current reaction (CC) is sensitive exclusively to electron-type
neutrinos, while the neutral current reaction (NC) is equally sensitive to all active
neutrino flavors ($x=e, \mu, \tau$).  The elastic scattering reaction (ES)
is sensitive to all flavors as well, but with reduced sensitivity to
$\nu_{\mu}$ and $\nu_{\tau}$.  Sensitivity to these three reactions allows
SNO to determine the electron and non-electron active neutrino components of the
solar flux~\cite{chen}.
The CC and ES reaction results have recently been presented~\cite{cc_prl}.
This Letter presents the first NC results and updated CC and ES results from SNO.

SNO~\cite{sno_nim} is a water Cherenkov detector located at a depth of 6010 m 
of water equivalent in the INCO, Ltd. Creighton mine near Sudbury, Ontario, Canada.  
The detector uses ultra-pure heavy water contained in a transparent acrylic spherical shell 12 m in
diameter to detect solar neutrinos.  Cherenkov photons generated in the heavy water are detected by 9456 photomultiplier tubes (PMTs) mounted on a stainless steel geodesic sphere 17.8 m in diameter.  The geodesic sphere is immersed in ultra-pure light water to provide shielding from radioactivity in both the PMT array and the cavity rock.

The data reported here were recorded between Nov. 2, 1999 and May 28, 2001
and represent a total of 306.4 live days, spanning the entire first phase of the experiment, in which
only D$_2$O was present in the sensitive volume.
The analysis procedure was similar to that described in~\cite{cc_prl}.
PMT times and hit patterns were used to reconstruct event vertices and directions
and to assign to each event a most probable kinetic energy, $T_{\rm eff}$.
The total flux of active ${}^{8}$B solar neutrinos with energies greater than 2.2 MeV (the NC reaction threshold) was measured with the NC signal (Cherenkov photons resulting from the 6.25 MeV $\gamma$ ray from neutron capture on deuterium.)
The analysis threshold was $T_{\rm eff}$$\geq 5$ MeV, providing sensitivity
to neutrons from the NC reaction.
Above this energy threshold, there were contributions from CC events in the 
D$_2$O, ES events in the D$_2$O and H$_2$O, capture of neutrons (both from the NC reaction
and backgrounds), and low energy Cherenkov background events.  

A fiducial volume was defined to only accept events which had reconstructed vertices within $550$~cm from the detector center
to reduce external backgrounds and systematic uncertainties associated with optics and
event reconstruction near the acrylic vessel.
The neutron response and systematic uncertainty was calibrated with a ${}^{252}$Cf source.  The deduced efficiency for neutron
captures on deuterium is  $29.9 \pm 1.1\%$ for a uniform source of neutrons in the D$_2$O.
The neutron detection efficiency within the fiducial volume and above the energy threshold 
is 14.4\%.
The energy calibration was updated from~\cite{cc_prl} with the ${}^{16}$N
calibration source~\cite{n16_nim} data and Monte Carlo calculations.  The energy response for electrons, updated for the lower
analysis threshold, was characterized as a Gaussian function with resolution $\sigma_{T} = -0.0684 + 0.331 \sqrt{T_{e}}+0.0425 T_{e}$, where $T_{e}$ is the true electron kinetic energy in MeV.  The energy scale uncertainty is $1.2\%$.

The primary backgrounds to the NC signal are due to low levels of uranium and thorium decay chain daughters (${}^{214}$Bi and ${}^{208}$Tl) in
the detector materials.   These activities generate free neutrons in the D$_2$O, from deuteron photodisintegration (pd), and low energy Cherenkov events.
{\textit{Ex-situ}} assays and {\textit{in-situ}} analysis of the low energy ($4-4.5$ MeV) Cherenkov
signal region provide independent uranium and thorium photodisintegration background measurements.
 
\begin{figure}
\includegraphics[width=3.4in]{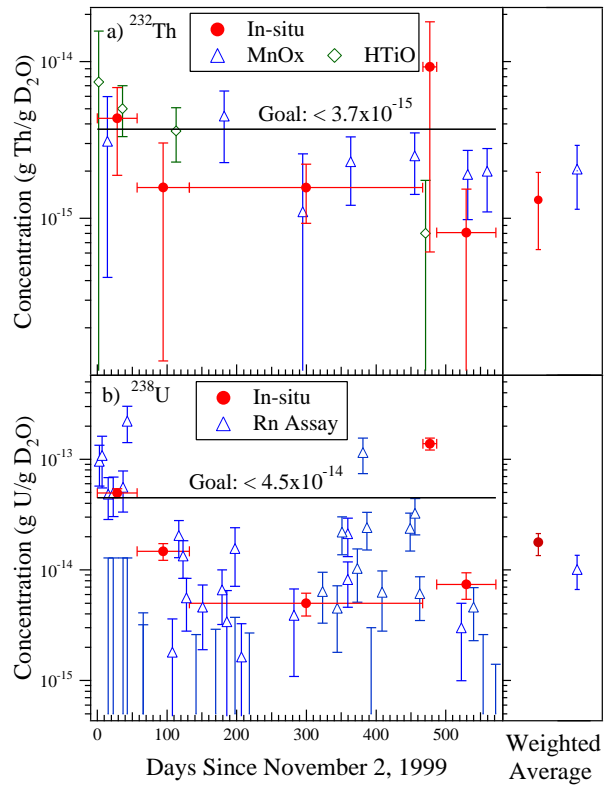}
\caption{\label{fig_pdback}Thorium (a) and uranium (b) backgrounds (equivalent equilibrium concentrations) in the D$_2$O deduced by {\textit{in situ}} and {\textit{ex situ}} techniques.  The MnO$_{\rm{x}}$ and HTiO radiochemical assay results, the Rn assay results,  and the {\textit{in situ}} Cherenkov signal determination of the backgrounds  are presented for the period of this analysis on the left-hand side of frames (a) and (b).  The right-hand side shows time-integrated averages including an additional sampling systematic uncertainty for the {\textit{ex situ}} measurement. }
\end{figure}

Two {\textit{ex situ}}  assay techniques were employed to determine average levels of uranium and thorium in water.  Radium ions were directly extracted from the water onto either MnO$_{\rm{x}}$ or hydrous Ti oxide (HTiO) ion exchange media.  Radon daughters in the U and Th chains were subsequently released, identified by $\alpha$ spectroscopy, or the radium was concentrated and the number of decay daughter $\beta$-$\alpha$ coincidences determined.  Typical assays circulated approximately 400 tonnes
of water through the extraction media.  These techniques provide isotopic identification of the decay daughters and contamination levels in the assayed water volumes, presented in Fig.~\ref{fig_pdback}~(a).  Secular equilibrium in the U decay chain was broken by the ingress of long-lived (3.8 day half-life) ${}^{222}$Rn in the experiment.  Measurements of this background were made by
periodically extracting and cryogenically concentrating ${}^{222}$Rn from water degassers.  Radon from several tonne assays was subsequently counted in ZnS(Ag) scintillation cells~\cite{manking}.  The Radon results are presented (as mass fractions in g(U)/g(D$_2$O)) in Fig.~\ref{fig_pdback}(b).

Independent measurements of U and Th decay chains were made by analyzing Cherenkov light produced by the radioactive decays.  The $\beta$ and $\beta$-$\gamma$ decays from the U and Th chains dominate the low energy monitoring window.  Events in this window monitor $\gamma$ rays that produce photodisintegration in these chains (E$_{\gamma}$~$>$~2.2 MeV).
Cherenkov events fitted within 450 cm from the detector center and extracted from the neutrino data set provide a time-integrated measure of these backgrounds over the same time period and within the fiducial volume of the neutrino analysis.  Statistical separation of {\textit{in situ}} Tl and Bi events was obtained by analyzing the Cherenkov signal isotropy.  
Tl decays always result in a $\beta$ and a 2.614 MeV $\gamma$, while in this energy window Bi decays are dominated by decays with only a $\beta$, and produce, on average, more anisotropic hit patterns.

Results from the {\textit{ex situ}} and {\textit{in situ}} methods are consistent with each other as shown on the right hand side of Figs.~\ref{fig_pdback}(a)~and~\ref{fig_pdback}(b).
For the ${}^{232}$Th chain, the weighted mean (including additional sampling systematic uncertainty) of the two determinations was used for the analysis.  The
${}^{238}$U chain activity is dominated by Rn ingress which is highly time dependent.  Therefore the {\textit{in-situ}}
determination was used for this activity as it provides the appropriate time weighting. 
The average rate of background neutron production from activities in the D$_2$O region is  
$1.0 \pm 0.2$ neutrons per day, leading to $44^{+8}_{-9}$ detected background events.
The production rate from external activities is $1.3^{+0.4}_{-0.5}$ neutrons per day, which leads to $27 \pm 8$ background events
since the neutron capture efficiency is reduced for neutrons born near the heavy water boundary.
  The total photodisintegration background corresponds to approximately $12\%$ of the number of NC neutrons predicted by the standard solar model from ${}^{8}$B neutrinos. 
 
Low energy backgrounds from Cherenkov events in the signal region were evaluated by using 
acrylic encapsulated sources  of U and Th deployed throughout the detector volume
and by Monte Carlo calculations.  Probability density functions (pdfs) in reconstructed vertex radius
derived from U and Th calibration data were used to determine the number of background Cherenkov
events from external regions which either entered or mis-reconstructed into the fiducial volume.
Cherenkov event backgrounds from activities in the D$_2$O were evaluated with Monte Carlo calculations.

Table~\ref{letable} shows the number of photodisintegration and Cherenkov background events (including
systematic uncertainties) due to activity
in the D$_2$O (internal region), acrylic vessel (AV), H$_2$O (external region), and PMT array.
Other sources of free neutrons in the D$_2$O region are cosmic ray events and atmospheric neutrinos.  To reduce these backgrounds, an additional neutron background cut imposed a 250-ms deadtime (in software) following every event in which the total number of PMTs which registered a hit was greater than 60. 
The number of remaining NC atmospheric neutrino events and background events generated by sub-Cherenkov
threshold muons is estimated to be small, as shown in Table~\ref{letable}.

\begingroup
\begin{table}
\squeezetable
\caption{\label{letable}Neutron and Cherenkov background events.}
\begin{ruledtabular}
\begin{tabular}{ll}
Source                                  & Events                        \\ \hline
D$_2$O photodisintegration              &  $44^{+8}_{-9}$               \\
H$_2$O + AV  photodisintegration        &  $27^{+8}_{-8}$               \\
Atmospheric $\nu$'s     and             &                               \\
sub-Cherenkov threshold $\mu$'s         &  $ 4 \pm 1$                   \\
Fission                                 &  $\ll1$                       \\
${}^{2}$H($\alpha,\alpha$)pn            &  $2 \pm 0.4$                  \\
${}^{17}$O($\alpha$,n)                  &  $\ll1$                       \\
Terrestrial and reactor $\bar{\nu}$'s   &  $1^{+3}_{-1}$                \\
External neutrons                       &  $\ll1$                       \\ \hline
Total neutron background                &  $78 \pm 12$                  \\ \hline
D$_2$O Cherenkov                        &  $20^{+13}_{\phantom{1}-6}$   \\
H$_2$O Cherenkov                        &  $3^{+4}_{-3}$                \\
AV Cherenkov                            &  $6^{+3}_{-6}$                \\
PMT Cherenkov                           &  $16^{+11}_{\phantom{1}-8}$   \\ \hline
Total Cherenkov background              &  $45^{+18}_{-12}$
\end{tabular}
\end{ruledtabular}
\end{table}
\endgroup

\begin{figure}
\psfrag{odot}{$\cos \theta_{\odot}$}
\includegraphics[width=3.73in]{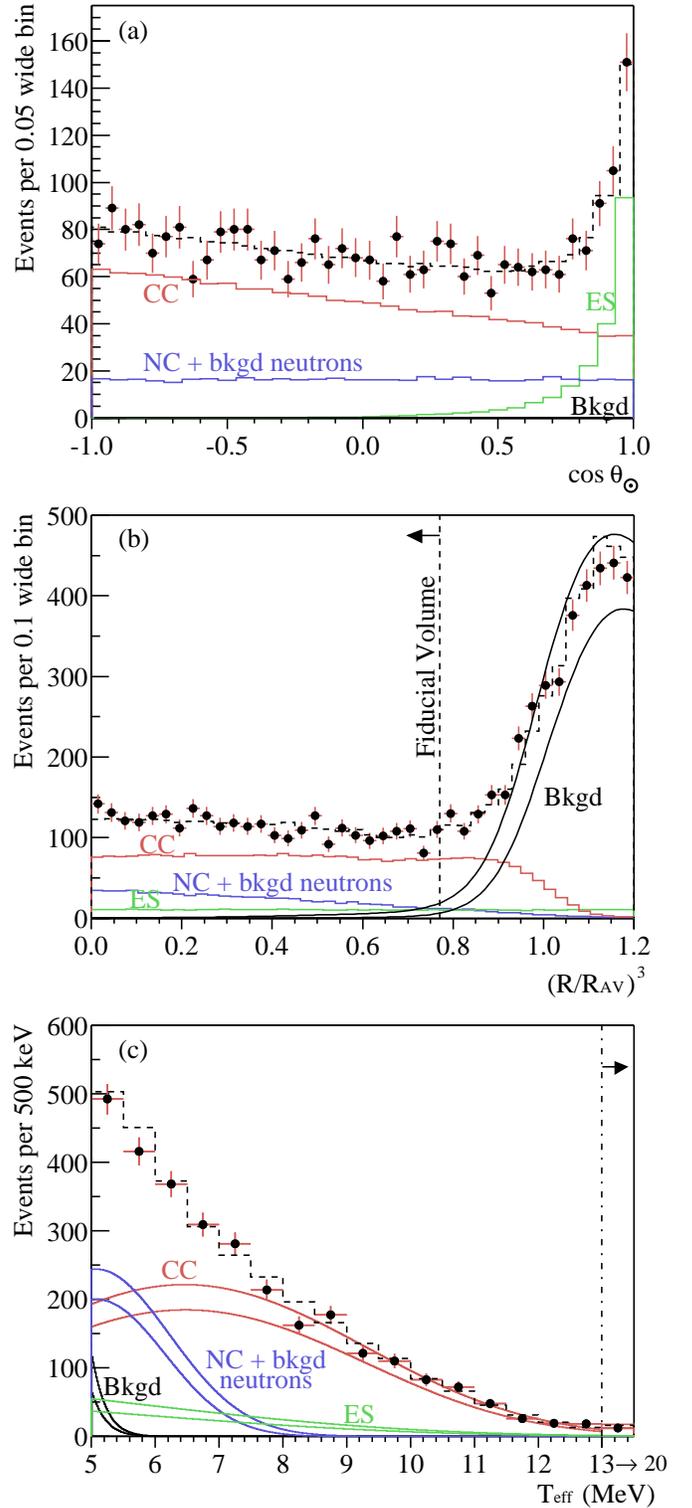}
\caption{\label{r3_comp}(a) Distribution of $\cos\theta_{\odot}$ for $R \le 550$ cm. (b) Distribution of the volume weighted radial variable $(R/R_{\rm AV})^{3}$.  (c) Kinetic energy for $R \le 550$ cm.  Also shown are the Monte Carlo predictions for CC, ES and NC + bkgd neutron events scaled to the fit results, and the calculated spectrum of Cherenkov background (Bkgd) events.  The dashed lines represent the summed components, and the bands show $\pm 1\sigma$ uncertainties.  All distributions are for events with $T_{\rm eff}$$\geq$5 MeV. }
\end{figure}

The data recorded during the pure D$_2$O detector phase are shown in Figure~\ref{r3_comp}.  These data have been analyzed using the same data reduction described in~\cite{cc_prl}, with the addition of the new neutron background cut, yielding 2928 events in the energy region selected for analysis, 5 to 20 MeV.
Fig.~\ref{r3_comp}(a) shows the  distribution of selected events in
the cosine of the angle between the Cherenkov event direction and the direction
from the sun ($\cos \theta_{\odot}$) for the
analysis threshold of $T_{\rm eff}$$\geq 5$ MeV and fiducial volume selection
of $R\le 550$ cm, where $R$ is the reconstructed event radius.
Fig.~\ref{r3_comp}(b) shows the distribution of events in the volume-weighted radial 
variable $(R/R_{\rm AV})^3$, where $R_{\rm AV}=600$~cm is the radius of the acrylic vessel.
Figure~\ref{r3_comp}(c) shows the kinetic energy spectrum of the selected events.

In order to test the null hypothesis, the assumption that there are only electron neutrinos in the solar
neutrino flux, the data are resolved into contributions from CC, ES, and NC
events above threshold using pdfs in $T_{\text{eff}}$, $\cos\theta_\odot$, and $(R/R_{\rm AV})^3$,
derived from Monte Carlo calculations
generated assuming no flavor transformation and the standard $^8$B spectral shape~\cite{ortiz}. 
Background event pdfs are included in the analysis with fixed amplitudes
determined by the background calibration.  The
extended maximum likelihood method used in the signal decomposition yields
$\nccfit$ CC events, $\nesfit$ ES events, and
$\nncfit$  NC events~\footnote{We note that this rate of neutron events also leads to a lower bound on the
        proton lifetime for ``invisible'' modes (based on the free neutron that would be left in deuterium (V.I. Tretyak and Yu.G. Zdesenko, Phys. Lett.{ \bf B505}, 59 (2001) ) in excess of $10^{28}$ years, approximately 3 orders
        of magnitude more restrictive than previous limits (J. Evans and R. Steinberg, Science,{ \bf 197}, 989 (1977).)
        The possible contribution of this mechanism to the solar neutrino NC background
	is ignored.}, where only statistical uncertainties are given. 
Systematic uncertainties on fluxes derived by repeating the signal decomposition
with perturbed pdfs (constrained by calibration data) are shown in Table~\ref{errors}.

\begingroup
\squeezetable
\begin{table}
\caption{\label{errors}Systematic uncertainties on fluxes. The experimental uncertainty for ES (not shown) is -4.8,+5.0 percent. $\dagger$ denotes CC vs NC anti-correlation.}
\begin{ruledtabular}
\begin{tabular}{llllll}
Source        &  CC Uncert.          & NC Uncert.         & $\phi_{\mu\tau}$ Uncert.    \\
                    & (percent)         &(percent)          &(percent)                  \\ \hline
Energy scale $\dagger$  & -4.2,+4.3 &-6.2,+6.1 &-10.4,+10.3  \\ 
Energy resolution $\dagger$  & -0.9,+0.0 &-0.0,+4.4 &-0.0,+6.8  \\ 
Energy non-linearity $\dagger$  & $\pm 0.1$ &  $\pm 0.4$ &  $\pm 0.6$    \\ 
Vertex resolution $\dagger$  & $\pm 0.0$ &  $\pm 0.1$ &  $\pm 0.2$    \\ 
Vertex accuracy & -2.8,+2.9 &$\pm 1.8$ &  $\pm 1.4$    \\ 
Angular resolution & -0.2,+0.2  &  -0.3,+0.3 &  -0.3,+0.3 \\ 
Internal source pd $\dagger$  & $\pm 0.0$ &  -1.5,+1.6 &-2.0,+2.2  \\ 
External source pd & $\pm 0.1$ &  -1.0,+1.0 &$\pm 1.4$    \\ 
D$_2$O Cherenkov $\dagger$  & -0.1,+0.2 &-2.6,+1.2 &-3.7,+1.7  \\ 
H$_2$O Cherenkov & $\pm 0.0$ &  -0.2,+0.4 &-0.2,+0.6  \\ 
AV Cherenkov & $\pm 0.0$ &  -0.2,+0.2 &-0.3,+0.3  \\ 
PMT Cherenkov $\dagger$  & $\pm 0.1$ &  -2.1,+1.6 &-3.0,+2.2  \\ 
Neutron capture & $\pm 0.0$ &  -4.0,+3.6 &-5.8,+5.2  \\ 
Cut acceptance & -0.2,+0.4 &-0.2,+0.4 &-0.2,+0.4  \\ \hline 
Experimental uncertainty & -5.2,+5.2 &  -8.5,+9.1 &  -13.2,+14.1 \\ \hline 
Cross section~\cite{crosssection} & $\pm 1.8$  & $\pm 1.3 $& $\pm 1.4$ 
\end{tabular}
\end{ruledtabular}
\end{table}
\endgroup

Normalized to the integrated rates above the kinetic energy threshold
of $T_{\rm eff}$$\geq 5$~MeV, the  flux of $^8$B neutrinos measured
with each reaction in SNO, 
assuming the standard spectrum shape~\cite{ortiz} is (all fluxes are presented in
units of $10^6~{\rm cm}^{-2} {\rm s}^{-1}$):
\nobreak{
\begin{eqnarray*}
\phi^{\text{SNO}}_{\text{CC}} & = & \snoccfluxshort \\
\phi^{\text{SNO}}_{\text{ES}} & = & \snoesfluxshort \\
\phi^{\text{SNO}}_{\text{NC}} & = & \snoncfluxshort. 
\end{eqnarray*}}
\noindent Electron neutrino cross sections are used to calculate all fluxes.
The CC and ES results reported here are consistent with the earlier SNO results~\cite{cc_prl} for $T_{\rm eff}$$\geq$6.75 MeV.  The excess of the NC flux over the CC and ES fluxes implies neutrino flavor transformations.

\begin{figure}
\includegraphics[width=3.7in]{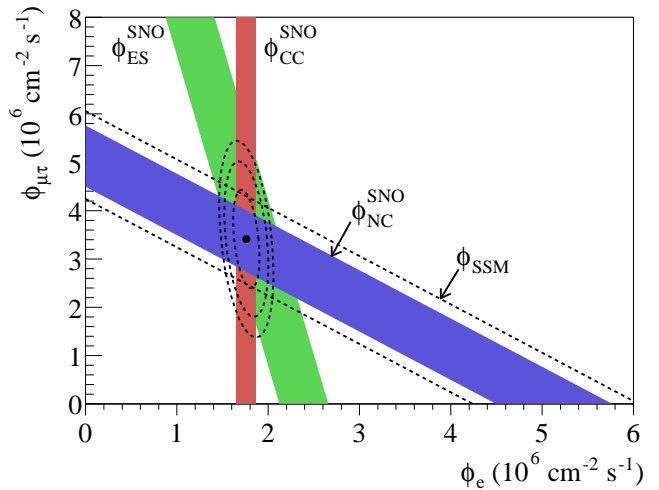}
\caption{\label{hime_plot}Flux of ${}^{8}$B solar neutrinos which are $\mu$ or $\tau$ flavor vs flux of electron neutrinos deduced from the three neutrino reactions in SNO.  The diagonal bands show the total ${}^{8}$B flux as predicted by the SSM~\cite{bp2000} (dashed lines) and that measured with the NC reaction in SNO (solid band).  The intercepts of these bands with the axes represent the $\pm 1\sigma$ errors.  The bands intersect at the fit values for $\phi_{e}$ and $\phi_{\mu\tau}$, indicating that the combined flux results are consistent with neutrino flavor transformation assuming no distortion in the ${}^{8}$B neutrino energy spectrum.}
\end{figure}

A simple change of variables resolves the data directly into electron ($\phi_{e}$) and non-electron ($\phi_{\mu\tau}$) components~\footnote{This change of variables allows a direct test of the null hypothesis of no flavor transformation ($\phi_{\mu\tau}=0$) without requiring calculation of the CC, ES, and NC signal correlations.},  
\begin{eqnarray*}
\phi_{e} & = & \snoeflux \\
\phi_{\mu\tau} & = & \snomutauflux 
\end{eqnarray*}
\noindent assuming the standard ${}^{8}$B shape.
Combining the statistical and systematic uncertainties in quadrature,
 $\phi_{\mu\tau}$ is $\snomutaufluxcomb$,
which is \nsigmassno$\sigma$  above zero, providing  strong
evidence for flavor transformation consistent with neutrino oscillations~\cite{maki,pontecorvo}.  Adding the Super-Kamiokande ES measurement of the ${}^{8}$B flux~\cite{superk} $\phi^{\text{SK}}_{\text{ES}}=\phisk$ as an additional constraint, we find $\phi_{\mu\tau}=\snomutaufluxsk$, which is \nsigmassk$\sigma$ above zero.
Figure~\ref{hime_plot} shows the flux of non-electron flavor active neutrinos
vs the flux of electron neutrinos deduced from the SNO data.  The three bands represent the one standard deviation measurements of the CC, ES, and NC rates. The error ellipses represent  the 68\%, 95\%, and 99\% joint probability contours for $\phi_{e}$ and $\phi_{\mu\tau}$.

Removing the constraint that the solar neutrino energy spectrum is undistorted, the signal decomposition
is repeated using only the $\cos \theta_{\odot}$ and $(R/R_{\rm AV})^3$ information.
The total flux of active ${}^{8}$B neutrinos measured with the NC reaction is 
\begin{equation*}
\phi_{\text{NC}}^{\text{SNO}} =  \snoncfluxunc 
\end{equation*}
\noindent which is in agreement with the shape constrained value above and with the standard solar model prediction~\cite{bp2000} for ${}^{8}$B, $\phi_{\text{SSM}} = \ssmflux.$ 

In summary, the results presented here are the first direct measurement
of the total flux of active $^8$B neutrinos arriving from the sun and provide strong
evidence for neutrino flavor transformation.  The CC and ES reaction rates are
consistent with the earlier results~\cite{cc_prl} and with the NC reaction rate under the hypothesis
of flavor transformation.  The total flux of ${}^{8}$B neutrinos measured with the NC reaction is in agreement with
the SSM prediction.

This research was supported by:  Canada: NSERC, Industry Canada, NRC,
Northern Ontario Heritage Fund Corporation, Inco, AECL, Ontario Power
Generation; US: Dept. of Energy; UK: PPARC. We thank the SNO technical
staff for their strong contributions.
\bibliography{nc_prl}
\end{document}